\documentclass[]{spie}  

\usepackage{amsmath,amsfonts,amssymb}
\usepackage{graphicx}
\usepackage[colorlinks=true, allcolors=blue]{hyperref}

\title{Quality Enhancement of Radiographic X-ray Images by Interpretable Mapping}

\author[a]{Hongxu Yang}
\author[b]{Najib Akram Maheen Aboobacker}
\author[b]{Xiaomeng Dong}
\author[b]{German Guillermo Vera Gonzalez}
\author[c]{Lehel Mihály Ferenczi}
\author[b]{Gopal Biligeri Avinash}
\affil[a]{GE Healthcare, Netherlands}
\affil[b]{GE Healthcare, USA}
\affil[c]{GE Healthcare, Hungary}
\authorinfo{Corresponding author: Hongxu Yang (hongxu.yang@gehealthcare.com)}

\begin{document} 
\maketitle

\begin{abstract}
X-ray imaging is the most widely used medical imaging modality. However, in the common practice, inconsistency in the initial presentation of X-ray images is a common complaint by radiologists. Different patient positions, patient habitus and scanning protocols can lead to differences in image presentations, e.g., differences in brightness and contrast globally or regionally. To compensate for this, additional work will be executed by clinical experts to adjust the images to the desired presentation, which can be time-consuming. Existing deep-learning-based end-to-end solutions can automatically correct images with promising performances. Nevertheless, these methods are hard to be interpreted and difficult to be understood by clinical experts. In this manuscript, a novel interpretable mapping method by deep learning is proposed, which automatically enhances the image brightness and contrast globally and locally. Meanwhile, because the model is inspired by the workflow of the brightness and contrast manipulation, it can provide interpretable pixel maps for explaining the motivation of image enhancement. The experiment on the clinical datasets show the proposed method can provide consistent brightness and contrast correction on X-ray images with accuracy of 24.75 dB PSNR and 0.8431 SSIM.
\end{abstract}


\section{INTRODUCTION}
\begin{figure} [htbp]
\begin{center}
\begin{tabular}{c} 
\includegraphics[width=16cm]{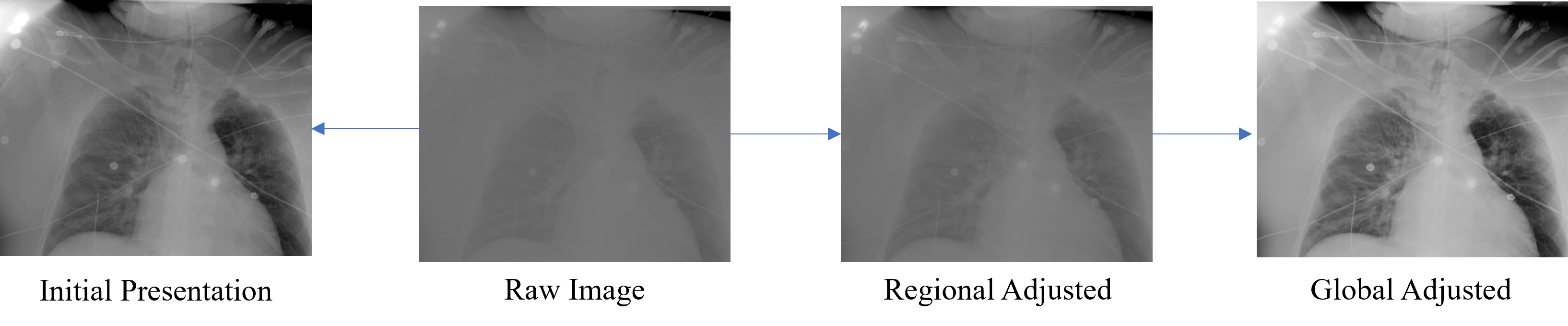}
\end{tabular}
\end{center}
\caption[Fig1]
{From raw image to left: global brightness and contrast adjustment for initial presentation after several iterations. From raw image to right: workflow for brightness and contrast enhancement at regional and global levels.}
\end{figure}

Radiographic X-ray projection images are widely considered because they are easy to use and provide detailed regional structure information. Nevertheless, due to various scanning positions and anatomies, X-ray images may have different initial presentations, which need extra tuning and efforts by technologists and/or radiologists when reviewing them~\cite{aboobacker2021improving}. Therefore, to better visualize the X-ray images, additional adjustments of the brightness and contrast (BC) in regional (also known as tissue-equalization, TE) and global levels are needed, as an example shown in Figure 1. Traditionally, complex image processing chains are designed with multi-level operations to obtain the image appearance as shown in Figure 1, including segmentation, pixel intensity analysis, Look-up-table (LUT) constructions, etc.~\cite{hoeppner2002equalized,sprawls2014optimizing} Although this conventional image processing chain provides a good interpretable solution for radiologist and engineers, it may require dedicated system designs for different anatomies, scanning positions and patient sizes. Despite the complexity of the system, additional user tuning is practically inevitable due to the use of pre-defined configurations, which cannot be automatically and adaptively adjusted based on observed images~\cite{aboobacker2021improving}. To tackle the above challenges, we propose a deep-learning-driven automated image brightness\&contrast enhancement/correction model, which is inspired by the multi-level image processing workflow~\cite{aboobacker2021improving,hoeppner2002equalized,sprawls2014optimizing}. 

\section{METHODS}
The method of automated X-ray image enhancement method is shown in Figure 2 (a), which consists of two key components: (1) pixel-wise parameter maps prediction of the input image, and (2) parameter maps transformation based on image processing inspired formulation mapping and LUTs. The input full resolution X-ray image is downsampled by a factor of 8 to fit the GPU memory, which is then processed by a ResUNet model~\cite{ronneberger2015u,he2016identity}. Specifically, the considered model has 5-level resolutions, for each resolution of network, a simple residual convolution with three conv layers is applied~\cite{he2016identity}. For each convolution operation, number of features are fixed to 32, which is followed by InstanceNorm and LeakyReLU. The final output of the ResUNet are interpretable parameter maps of regional and global LUT functions for each pixel. Finally, the predicted maps are upsampled back to full resolution for formula remap projections, which is a pixel-level operation by using the input image as the input. With the obtain outputs, gray scale intensities are finally rescaled back to [0, 65535]. This method provides an interpretable parametrical map for each pixel to construct a pixel-wise LUT, as shown in Figure 2 (b), which is different to the conventional image processing methods where only image-level LUTs are provided~\cite{aboobacker2021improving,hoeppner2002equalized,sprawls2014optimizing}. Therefore, with aiding of deep learning, the pixel-wise LUTs can adaptively adjust the image presentation regionally and globally based on observed anatomies, scanning positions and patient sizes.
\begin{figure} [htbp]
\begin{center}
\begin{tabular}{c} 
\includegraphics[width=16cm]{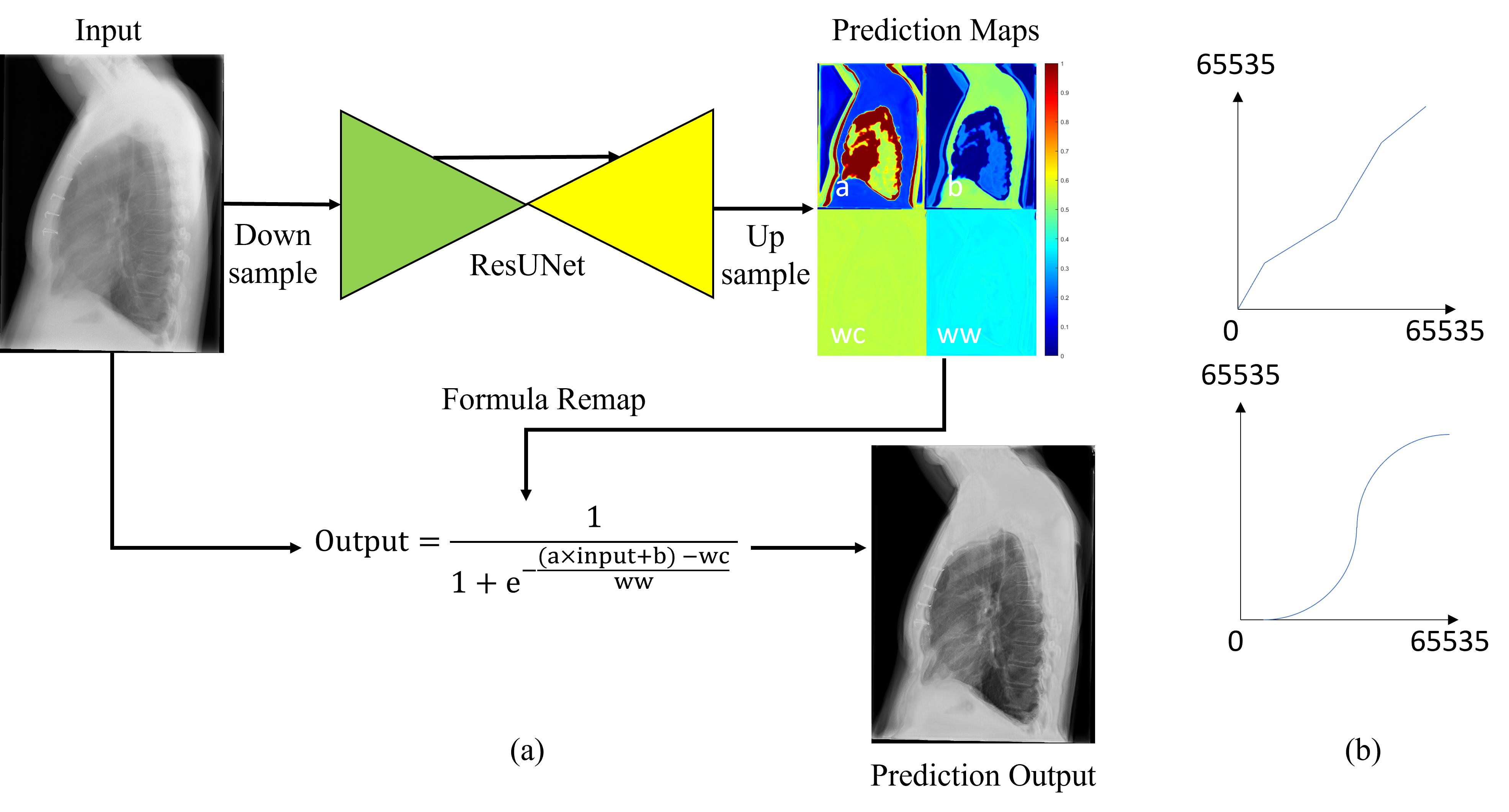}
\end{tabular}
\end{center}
\caption[Fig2]
{(a) Interpretable maps prediction and enhanced image generation based on input image, predicted maps (a, b, wc and ww) and remap formula. (b) Regional LUT and global LUT for dynamic range of [0,65535].}
\end{figure}

To train the model of the ability to predict parameter maps, a specific loss function is designed based on LUTs parameters and remapping formula above. Specifically, the local adjustment is formulated by changing the pixel intensities for different dynamic range, as shown in top of Figure 2 (b), where the LUT is constructed by a piecewise monotonically increasing linear function. Therefore, the parameters of the function are pixel-wise slope \textbf{a}, and bias \textbf{b}. In contrast, the global adjustment is form by window-center (\textbf{wc}) and window-width (\textbf{ww}) of conventional brightness and contrast adjustment with sigmoid like function. The objective function is formulated in Eqn. 1 based on mean square error (MSE) at pixel level:
\begin{equation}
\small
Loss=MSE(img_{pred},img_{GT})+MSE(a_{pred},a_{gt})+MSE(b_{pred},b_{gt})+MSE(ww_{pred},ww_{gt})+MSE(wc_{pred},wc_{gt})
\end{equation}

During the training, data augmentations are randomly performed on-the-fly, which includes random rotation, mirror, shear, etc. The parameters are learned by the above loss function using Adam optimizer with learning rate of 1e-5. The training is terminated after converged on validation dataset with mini-batch size equals to 1.
\section{RESULTS}
\begin{figure} [htbp]
\begin{center}
\begin{tabular}{c} 
\includegraphics[width=14cm]{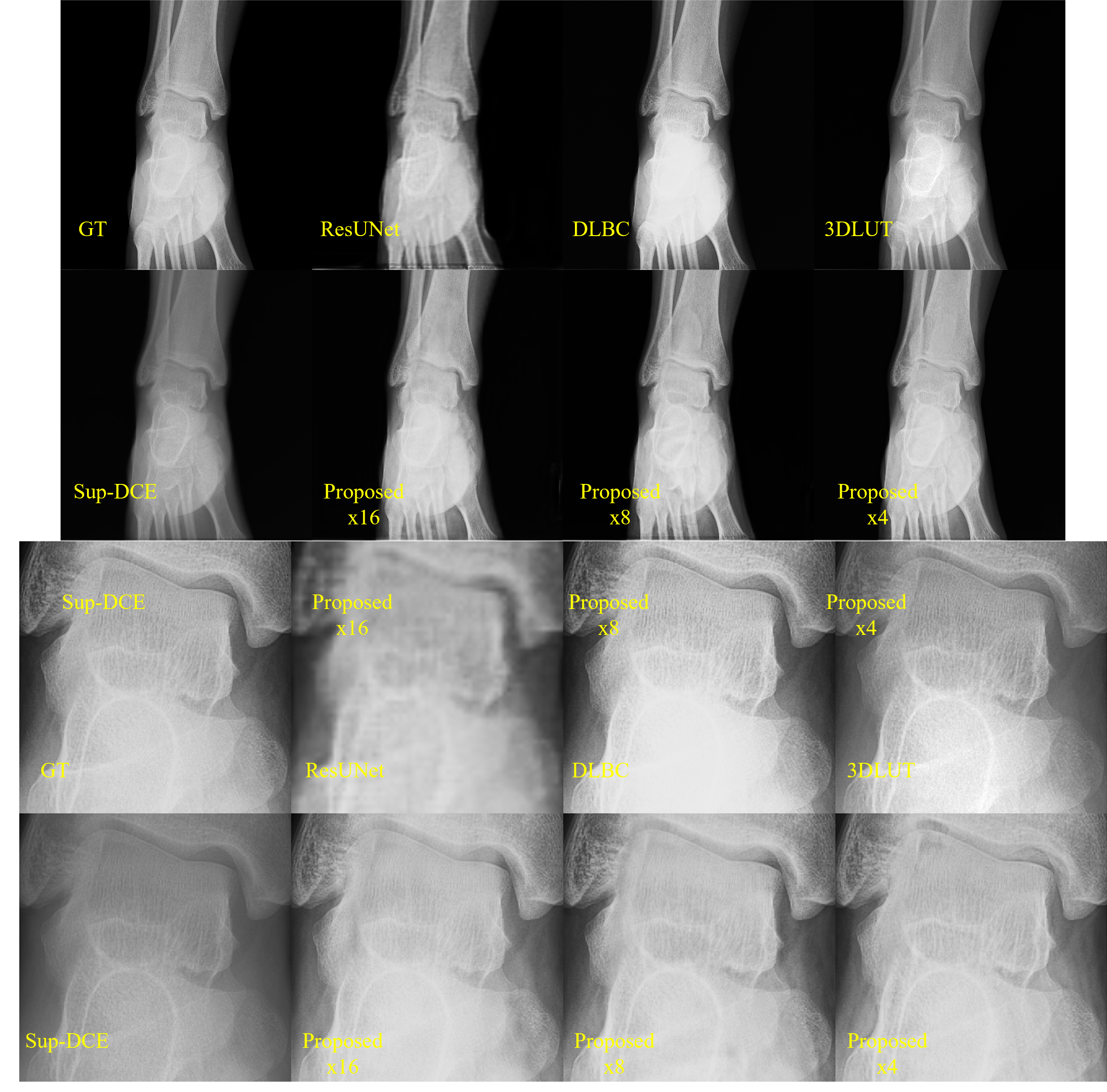}
\end{tabular}
\end{center}
\caption[Fig3]
{Top eight examples are selected from different models in Table 1, including the ground truth (GT). Bottom eight examples are enlarged ankle regions of corresponding model. The details of the bone structures are preserved by our method. }
\end{figure}

\begin{figure} [htbp]
\begin{center}
\begin{tabular}{c} 
\includegraphics[width=16cm]{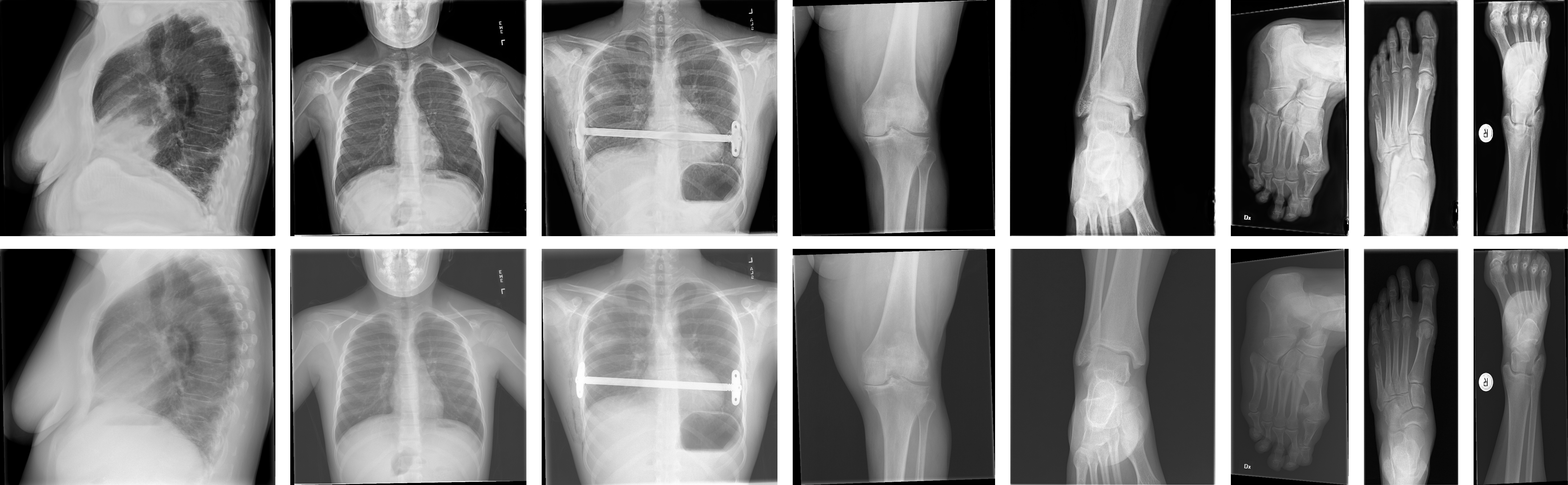}
\end{tabular}
\end{center}
\caption[Fig4]
{Top: Outputs from the proposed method, which are adjusted regionally and globally for brightness and contrast. Bottom: Automated brightness and contrast adjustment globally only.}
\end{figure}

To validate the proposed method, we used 429 X-ray images from clinical sites around the world. The dataset consists of different anatomy and scanning positions from daily clinical practice. We developed an in-house engineering tool to efficiently annotate the images with correct parameters, which provides consistent image look as the ground truth in Eqn. 1. The dataset is randomly partitioned to 356:30:43 for train/validation/test purposes. Specifically, due to different pre-processing of X-ray workflow and collimator size, X-ray image sizes are ranging from 1000 pixels to 4096 pixels for different anatomy. We compared image enhancement models of image-to-image translation by ResUNet at donwsample factor of 8, deep-learning-based brightness and contrast (DLBC) correction by directly predicting window-center and window-level (re-implemented based on proposal)~\cite{aboobacker2021improving}, 3DLUT by predicting intensity space weighted LUT~\cite{zeng2020learning} and supervised Zero-DCE, i.e. Sup-DCE~\cite{li2021learning}, as the references. Specifically, the 3DLUT consists of eight different LUT bases with dimension of 33 points~\cite{zeng2020learning}. As for the proposed method, downscale factor value of 4, 8, and 16 are considered for performance comparison. All the model trainings were performed on A100 GPU with 40GB RAM from a DGX box. The model performances are evaluated by peak signal-to-noise-ratio (PSNR) and structure similarity index measurement (SSIM) on the testing images.

\begin{table}[htbp]
\caption {Performance comparisons of different methods by PSNR and SSIM. The processing time is measured on A100.}
\centering  
\begin{tabular}{cccc}
\hline
Method& PSNR (dB) & SSIM & Time (sec.)\\ \hline
ResUNet (downscale=8) & 25.10 & 0.7652 & 0.0275 \\ \hline
DLBC & 22.97 & 0.7500 & 0.0127 \\ \hline
3DLUT & 24.33 & 0.8255 & 0.0127 \\ \hline
Sup-DCE& 19.59 & 0.7275 & 0.0208 \\ \hline
Proposed (downscale=16) & 24.36 & 0.8508 & 0.0259 \\ \hline
Proposed (downscale=8)& 24.75 & 0.8431 & 0.0262 \\ \hline
Proposed (downscale=4)& 24.89 & 0.8461 & 0.0281 \\ \hline
\end{tabular}  
\end{table} 

The performances of different models on testing dataset are summarized in Table 1. The Sup-DCE model has the worst performance, since it uses the simplest multi-level exponential projection, which cannot generate contrast enhancement with balanced presentation. ResUNet has the best performance in PSNR, however, by visually comparing the results, this method has the blurriest artifacts due to architecture design of up-sampling in the end, leading to low SSIM. The brightness and contrast enhancement algorithm (DLBC) cannot generate a good regional contrast, as it is designed for global correction. The 3DLUT method has close performance to the proposal. However, when comparing the regional details of the example images in Figure 3, it has worse adjustment of the local contrast, which is because LUTs cannot operate the image pixels for specific regions. 

The proposal has the best performance, meanwhile, different downscale factors have limited performance differences. This result indicates the low resolution of predicted parameter maps still can preserve the good image quality in full-resolution images. As for the consistency of the image presentation among different images, as shown in Figure 4, the proposed method consistent presentations compared to straightforward window-width and window-center manipulation on different image with different anatomies and positions in both global and region perspective.

\section{CONCLUSIONS}
We have proposed an interpretable parameter mapping method for radiographic X-ray image enhancement, which is the first study to decompose the conventional image processing workflow to end-to-end deep learning frameworks with interpretable remapping approach. With the proposed method, high-resolution X-ray images can be adaptively enhanced to have consistent presentation across different anatomies and scanning positions. Experimental validation shows the proposed method achieves promising results on the clinical datasets, which are better than the current methods via deep learning. 

Automated X-ray image enhancement with presentation consistency across different anatomies and scanning positions can reduce the image manipulation time of radiologist. Therefore, clinical experts can better focus on image itself for diagnosis with better concentration. The proposed method can automatically generate the consistent X-ray images with interpretable parameter maps, which helps clinical experts to better understand the behavior of the deep learning models. The experimental results show the proposed method achieved 24.75 dB PSNR and 0.8431 SSIM in full-resolution images. Based on the results, we will investigate the possible semi-supervised or unsupervised method to facilitate the development on large scale datasets.

\section*{STATEMENT} 
This work is original and in its present form has not been submitted elsewhere in any form. The research is internally funded by GE Healthcare.

\bibliography{main} 

\begin{thebibliography}{1}

\bibitem{aboobacker2021improving}
Aboobacker, N. A.~M., Gonzalez, G.~V., Zhang, F., Wanek, J., Xue, P., Rao, G., and Ye, D.~H., ``Improving presentation consistency of radiographic images using deep learning,'' in [{\em Medical Imaging 2021: Physics of Medical Imaging}{\nolinebreak\hspace{0.1em}]},   {\bf 11595},  636--643, SPIE (2021).

\bibitem{hoeppner2002equalized}
Hoeppner, S., Maack, I., Neitzel, U., and Stahl, M., ``Equalized contrast display processing for digital radiography,'' in [{\em Medical Imaging 2002: Visualization, Image-Guided Procedures, and Display}{\nolinebreak\hspace{0.1em}]},   {\bf 4681},  617--625, SPIE (2002).

\bibitem{sprawls2014optimizing}
Sprawls, P., ``Optimizing medical image contrast, detail and noise in the digital era,'' {\em Medical Physics International}~{\bf 2}(1) (2014).

\bibitem{ronneberger2015u}
Ronneberger, O., Fischer, P., and Brox, T., ``U-net: Convolutional networks for biomedical image segmentation,'' in [{\em Medical image computing and computer-assisted intervention--MICCAI 2015: 18th international conference, Munich, Germany, October 5-9, 2015, proceedings, part III 18}{\nolinebreak\hspace{0.1em}]},   234--241, Springer (2015).

\bibitem{he2016identity}
He, K., Zhang, X., Ren, S., and Sun, J., ``Identity mappings in deep residual networks,'' in [{\em Computer Vision--ECCV 2016: 14th European Conference, Amsterdam, The Netherlands, October 11--14, 2016, Proceedings, Part IV 14}{\nolinebreak\hspace{0.1em}]},   630--645, Springer (2016).

\bibitem{zeng2020learning}
Zeng, H., Cai, J., Li, L., Cao, Z., and Zhang, L., ``Learning image-adaptive 3d lookup tables for high performance photo enhancement in real-time,'' {\em IEEE Transactions on Pattern Analysis and Machine Intelligence}~{\bf 44}(4),  2058--2073 (2020).

\bibitem{li2021learning}
Li, C., Guo, C., and Loy, C.~C., ``Learning to enhance low-light image via zero-reference deep curve estimation,'' {\em IEEE transactions on pattern analysis and machine intelligence}~{\bf 44}(8),  4225--4238 (2021).

\end{thebibliography}
\bibliographystyle{spiebib} 

\end{document}